# Generalized molecular chaos hypothesis and the *H*-theorem: Problem of constraints and amendment of nonextensive statistical mechanics

Sumiyoshi Abe[1,2,3]

[1]*Department of Physical Engineering, Mie University, Mie 514-8507, Japan**)*
[2]*Institut Supérieur des Matériaux et Mécaniques Avancés, 44 F. A. Bartholdi, 72000 Le Mans, France*
[3]*Inspire Institute Inc., McLean, Virginia 22101, USA*

**Abstract**   Quite unexpectedly, kinetic theory is found to specify the correct definition of average value to be employed in nonextensive statistical mechanics. It is shown that the normal average is consistent with the generalized Stosszahlansatz (i.e., molecular chaos hypothesis) and the associated *H*-theorem, whereas the *q*-average widely used in the relevant literature is not. In the course of the analysis, the distributions with finite cut-off factors are rigorously treated. Accordingly, the formulation of nonextensive statistical mechanics is amended based on the normal average. In addition, the Shore-Johnson theorem, which supports the use of the *q*-average, is carefully reexamined, and it is found that one of the axioms may not be appropriate for systems to be treated within the framework of nonextensive statistical mechanics.

______________________________

*) Permanent address

## I. INTRODUCTION

There exist a number of physical systems that possess exotic properties in view of traditional Boltzmann-Gibbs statistical mechanics. They are said to be statistical-mechanically anomalous, since they often exhibit and realize broken ergodicity, strong correlation between elements, (multi)fractality of phase-space/configuration-space portraits, and long-range interactions, for example. In the past decade, nonextensive statistical mechanics [1,2], which is a generalization of the Boltzmann-Gibbs theory, has been drawing continuous attention as a possible theoretical framework for describing these systems.

The current prevailing formulation of nonextensive statistical mechanics [3] is based on *two simultaneous changes* of the ordinary maximum entropy principle [4]: one is a generalized form of entropy [5], which is given in terms of a probability distribution, $\{p_i\}_{i=1,2,...,W}$ (with $W$ being the number of accessible states) by

$$S_q[p] = \frac{1}{1-q}\left[\sum_{i=1}^{W}(p_i)^q - 1\right], \qquad (1)$$

and the other is the modified definition of average value [3] of a physical quantity, $Q = \{Q_i\}_{i=1,2,...,W}$,

$$<Q>_q = \frac{\sum_{i=1}^{W} Q_i (p_i)^q}{\sum_{j=1}^{W} (p_j)^q}. \qquad (2)$$

$S_q[p]$ and $<Q>_q$ are referred to as the Tsallis entropy with the entropic index $q\,(>0)$ and the $q$-average, respectively. They respectively tend to the Boltzmann-Gibbs-Shannon entropy, $S[p] = -\sum_{i=1}^{w} p_i \ln p_i$, and the normal average, $<Q> = \sum_{i=1}^{W} Q_i\, p_i$, in the limit $q \to 1$ (provided that the Boltzmann constant is set equal to unity). If $S_q[p]$ is maximized under the constraints on $<Q>_q$ as well as the normalization condition $\sum_{i=1}^{W} p_i = 1$, then the resulting stationary distribution reads [3]

$$\tilde{p}_i = \frac{1}{\tilde{Z}_q(\beta)} e_q\!\left(-(\beta/\tilde{c}_q)(Q_i - \tilde{Q}_q)\right), \tag{3}$$

$$\tilde{Z}_q(\beta) = \sum_{i=1}^{W} e_q\!\left(-(\beta/\tilde{c}_q)(Q_i - \tilde{Q}_q)\right). \tag{4}$$

Here, $\beta$ is the Lagrange multiplier associated with the constraint on the $q$-average, $\tilde{c}_q \equiv \sum_{i=1}^{W} (\tilde{p}_i)^q$, and $\tilde{Q}_q$ is the $q$-average of $Q$ with respect to $\tilde{p}_i$. $e_q(x)$ is termed the $q$-exponential function defined by

$$e_q(x) = [1 + (1-q)x]_+^{1/(1-q)} \tag{5}$$

with the notation $[a]_+ \equiv \max\{0, a\}$, which is the inverse function of the $q$-logarithmic function

$$\ln_q(x) = \frac{1}{1-q}\left(x^{1-q} - 1\right). \tag{6}$$

They converge to the ordinary $e^x$ and $\ln x$ in the limit $q \to 1$, respectively. The distribution in Eq. (3) has received much attention, since it is a power-law distribution of the Zipf-Mandelbrot type for $q > 1$, which is often observed in nature.

For our later discussion, it is convenient to rewrite Eq. (3) as follows:

$$\tilde{p}_i = \tilde{N}(\tilde{\lambda}) \, e_q(-\tilde{\lambda} \, Q_i), \tag{7}$$

where

$$\tilde{N}^{-1}(\tilde{\lambda}) = \sum_{i=1}^{W} e_q(-\tilde{\lambda} \, Q_i), \tag{8}$$

$$\tilde{\lambda} = \frac{(1-q)\beta}{\tilde{c}_q + (1-q)\beta \tilde{Q}_q}. \tag{9}$$

The distribution in Eq. (3) [or (7)] is refereed to here as the "$e_q$-distribution".

Although a lot of efforts have been devoted to understanding the physical meaning of the Tsallis entropy in the literature (see, for example, Refs. [6-8] and the references therein), less attention has been paid to the concept of the $q$-average. Recently, it has been shown [9] that Eq. (2) is unstable under small deformations of the probability distribution, in general, unless the Boltzmann-Gibbs limit, $q \to 1$. This discovery naturally requires one to carefully examine the $q$-average formalism proposed in Ref. [3]. In other words, one should reconsider the normal average

$$<Q> = \sum_{i=1}^{W} Q_i \, p_i \qquad (10)$$

for nonextensive statistical mechanics. To determine which the correct definition is, Eq. (2) or Eq. (10), some physical principles are needed.

In this paper, we report an unexpected result that kinetic theory selects the normal-average formalism. This is indeed unexpected since usually the discussion of a kinetic equation itself is not directly concerned with average value of any physical quantity. A generalized Stosszahlansatz (i.e., molecular chaos hypothesis) and the associated $H$-theorem for the Tsallis-type $H$-function are shown to lead to the conclusion that the normal-average formalism is consistent, whereas the $q$-average formalism is not. In this analysis, the distributions with finite cut-off factors are rigorously treated. Furthermore, the Shore-Johnson theorem, which is supportive for the use of the $q$-average, is carefully reexamined, and it is found that one of the five axioms of Shore and Johnson is not physically appropriate for systems to be treated by nonextensive statistical mechanics.

This paper is organized as follows. In Sec. II, the normal-average formalism for nonextensive statistical mechanics is revisited. In Sec. III, a generalized Stosszahlansatz and the associated $H$-theorem are discussed for the Tsallis-type $H$-function. There, it is shown that the normal-average formalism is consistent, whereas the $q$-average formalism is not. In Sec. IV, the Shore-Johnson theorem [10], which is known to support the use of the *q*-average, is reexamined from the physical viewpoint. Section V contains concluding remarks.

## II.  NORMAL-AVERAGE FORMALISM

This section is devoted to formulating the maximum Tsallis-entropy method with normal averages, which turns out to be an amendment of nonextensive statistical mechanics.

Actually, such a discussion has already been made in an incomplete form in Ref. [5], and the complete formulation has been presented in Refs. [11,12], where the Shore-Johnson theorem is shown to support the use of *q*-averages (see Sec. IV). Since the normal-average formalism does not seem to be prevailing, it may be appropriate to recapitulate it here. In addition, the discussions in Refs. [11,12] will be developed further.

Nonextensive statistical mechanics with the normal average can be formulated by considering the following functional:

$$\Phi[p;\alpha,\beta] = S_q[p] - \alpha \left( \sum_{i=1}^{W} p_i - 1 \right) - \beta \left( \sum_{i=1}^{W} Q_i\, p_i - <Q> \right), \qquad (11)$$

where $\alpha$ and $\beta$ are the Lagrange multipliers. Maximization condition of this functional reads

$$\frac{\delta}{\delta p_i} \Phi[p; \alpha, \beta] = 0. \tag{12}$$

To eliminate $\alpha$, it is convenient to combine the normalization condition, $\sum_{i=1}^{W} p_i = 1$, with the following identical relation:

$$\sum_{i=1}^{W} p_i \frac{\delta}{\delta p_i} \Phi[p; \alpha, \beta] = 0. \tag{13}$$

The resulting stationary distribution is given by

$$\hat{p}_i = \frac{1}{\hat{Z}_q(\beta)} E_q\left(-(\beta / \hat{c}_q)(Q_i - \hat{Q})\right), \tag{14}$$

$$\hat{Z}_q(\beta) = \sum_{i=1}^{W} E_q\left(-(\beta / \hat{c}_q)(Q_i - \hat{Q})\right). \tag{15}$$

Here, $\hat{Q}$ is the normal average of $Q$ with respect to $\hat{p}_i$. The function, $E_q(x)$, appearing in these equation is defined by

$$E_q(x) = [1 + (1 - 1/q)x]_+^{1/(q-1)}, \tag{16}$$

whose inverse function is

$$\text{Ln}_q(x) = \frac{q}{q-1}\left(x^{q-1} - 1\right). \tag{17}$$

As in Eq. (7), we rewrite Eq. (15) in the following form:

$$\hat{p}_i = \hat{N}(\hat{\lambda}) \, E_q(-\hat{\lambda} \, Q_i), \qquad (18)$$

where

$$\hat{N}^{-1}(\hat{\lambda}) = \sum_{i=1}^{W} E_q(-\hat{\lambda} \, Q_i), \qquad (19)$$

$$\hat{\lambda} = \frac{(q-1)\beta}{q\,\hat{c}_q + (q-1)\beta\,\hat{Q}}. \qquad (20)$$

The distribution in Eq. (14) [or (18)] is refereed to as the "$E_q$-distribution".

It may be useful to present the following relations between the functions, $e_q(x)$, $E_q(x)$, $\ln_q(x)$, and $\mathrm{Ln}_q(x)$:

$$e_q(x) = E_{2-q}((2-q)x), \qquad E_q(x) = e_{2-q}(x/q), \qquad (21)$$

$$\ln_q(x) = -\frac{1}{q}\mathrm{Ln}_q(1/x), \qquad \mathrm{Ln}_q(x) = -q\,\ln_q(1/x). \qquad (22)$$

We wish to emphasize some important points. Firstly, comparing Eq. (3) with Eq. (14), the exponents obey the "duality": $1-q \leftrightarrow q-1$. Such an observation, however, holds only for the exponents. $\tilde{\lambda}$ in Eq. (9) and $\hat{\lambda}$ in Eq. (20) are radically different, and the values of $\tilde{Q}_q$ and $\hat{Q}$ cannot be related to each other by the replacement $1-q \leftrightarrow q-1$. Therefore, the thermodynamic quantities such as the specific heat and pressure take totally different values in the normal-average and *q*-average formalisms, when *Q* is taken to be the system Hamiltonian. Thus, *there exists no equivalence between these two formalisms: they are quite disparate each other*.

As already mentioned, we certainly need physical principles to specify the correct definition of average values to be employed in nonextensive statistical mechanics. In the next section, we present one such principle.

## III. GENERALIZED STOSSZAHLANSATZ AND ASSOCIATED *H*-THEOREM

Recently, a number of efforts [13-19] have been devoted to clarify if the stationary distribution in nonextensive statistical mechanics can be understood in terms of relaxation described by Boltzmann-like kinetic theory. There, Boltzmann's original Stosszahlansatz is generalized in order to include correlation between colliding particles. The results obtained so far are intriguing. If the Tsallis-type *H*-function [recall Eq. (1)]

$$H_q(\mathbf{r},t) = \frac{1}{1-q}\int d^3\mathbf{v}\left[f(\mathbf{r},\mathbf{v},t) - f^q(\mathbf{r},\mathbf{v},t)\right] \qquad (23)$$

is considered for the distribution of the position and velocity of a particle, $f(\mathbf{r},\mathbf{v},t)$, then the system relaxes to the stationary state described not by the $e_q$-distribution in Eq. (7) but by $E_q$-distribution in Eq. (18) [13,14]. Complimentarily, if $H_{2-q}(\mathbf{r},t)$ is used as a *H*-function, then $e_q$-distribution can be realized as the relaxed state [18]. This implies that, as long as based on the *q*-average formalism, the one and only consistent case is the limiting one $q \to 1$, i.e., the Boltzmann-Gibbs theory, showing that the correct definition of average is not the $q$-average.

In what follows, we show that the normal-average formalism is fully consistent with the generalized *H*-theorem for the *H*-function in Eq. (23). In particular, we shall develop a rigorously discussion about the distributions with compact supports (i.e., finite cut-off factors), a point of which is not investigated in the previous works [13-17].

A basic observation is as follows. The time derivative of $H_q(\mathbf{r},t)$ in Eq. (23) is given by

$$\frac{\partial H_q(\mathbf{r},t)}{\partial t} = \frac{1}{1-q} \int d^3\mathbf{v} \left[1 - q f^{q-1}(\mathbf{r},\mathbf{v},t)\right] \frac{\partial f(\mathbf{r},\mathbf{v},t)}{\partial t}. \tag{24}$$

The power appearing in the quantity in the integrand on the right-hand side is $q-1$. This fact strongly suggests that the relaxed state may not be the $e_q$-distribution but the $E_q$-distribution.

Let us suppose $f(\mathbf{r},\mathbf{v},t)$ to obey the equation of the following form:

$$\frac{\partial f}{\partial t} + \mathbf{v}\cdot\nabla f + \frac{\mathbf{F}}{m}\cdot\frac{\partial f}{\partial \mathbf{v}} = C(f), \tag{25}$$

where *m* is the mass of the particle and **F** is a force exerted on the particle that is assumed to be independent of the velocity for the sake of simplicity. $C(f)$ represents the collision term, which plays a central role in the subsequent discussion. As in the standard discussion, we here consider only the simplest binary collision satisfying symmetries and conservation laws, $(\mathbf{v}, \mathbf{v}_1) \rightarrow (\mathbf{v}', \mathbf{v}_1')$, with arbitrary $\mathbf{v}_1$. Write the collision term as follows:

$$C(f) = \int d\omega \, d^3\mathbf{v} \, \sigma \, V_r \, R(f, f_1; f', f_1'), \tag{26}$$

where $V_r$ is the magnitude of the relative velocity of two particles before collision, $\sigma$ the scattering cross section, and $\omega$ the solid angle appearing in geometry of collision kinematics. $R(f, f_1; f', f_1')$ describes the correlation difference in the system before and after collision in terms of the distributions, $f(\mathbf{r}, \mathbf{v}, t)$, $f_1 \equiv f(\mathbf{r}, \mathbf{v}_1, t)$, $f' = f(\mathbf{r}, \mathbf{v}', t)$, and $f_1' = f(\mathbf{r}, \mathbf{v}_1', t)$.

In the ordinary Stosszahlansatz, two colliding particles have no correlation, and therefore their joint distribution is factorized:

$$f^{(2)}(\mathbf{r}, \mathbf{r}_1, \mathbf{v}, \mathbf{v}_1, t) = f(\mathbf{r}, \mathbf{v}, t) f(\mathbf{r}_1, \mathbf{v}_1, t). \tag{27}$$

Accordingly, the correlation difference reads

$$R(f, f_1; f', f_1') = f' f_1' - f f_1. \tag{28}$$

In a nonextensive statistical mechanical system, the particles are always strongly correlated and the factorization of the joint distribution is not realized. Here, we consider a specific type of correlation that is suggested by the inherent mathematical structure of nonextensive statistical mechanics in the normal-average formalism. Let us introduce an operation:

$$x *_q y = [x^{q-1} + y^{q-1} - 1]_+^{1/(q-1)}. \tag{29}$$

This operation has the following properties:

$$E_q(x) *_q E_q(y) = E_q(x+y), \tag{30}$$

$$\mathrm{Ln}_q(x *_q y) = \mathrm{Ln}_q(x) + \mathrm{Ln}_q(y), \tag{31}$$

and $x *_q y$ tends to $xy$ in the limit $q \to 1$. It is related to the so-called $q$-product [21,22], $x \otimes_q y$, simply as follows:

$$x *_q y = x \otimes_{2-q} y. \tag{32}$$

Also, recall the definition, $[a]_+ \equiv \max\{0, a\}$. This is, however, not precise enough for our subsequent purpose, since we shall have to compare two infinitesimals, $[a]_+$ and $[b]_+$ for $a, b < 0$. A more precise meaning of $[a]_+ = 0$ for $a < 0$ may be

$$[a]_+ = 0 \equiv \lim_{s \to 0} a \exp\left(\frac{a}{s^2}\right) \qquad (a < 0). \tag{33}$$

Accordingly, it is natural to define the ratio as follows:

$$\frac{[b]_+}{[a]_+} \equiv \lim_{s \to 0} \frac{b}{a} \exp\left(\frac{b-a}{s^2}\right) \qquad (a, b < 0), \tag{34}$$

which implies that

$$\frac{[b]_+}{[a]_+} = \begin{cases} 0 & (b < a < 0) \\ \infty & (a < b < 0) \end{cases}. \tag{35}$$

This scheme plays an important role in dealing with distributions with compact supports.

Now, our proposal for generalizing the ordinary Stosszahlansatz is to replace Eq. (27) with

$$f_q^{(2)}(\mathbf{r}, \mathbf{r}_1, \mathbf{v}, \mathbf{v}_1, t) = K_q\, f(\mathbf{r}, \mathbf{v}, t) *_q f(\mathbf{r}_1, \mathbf{v}_1, t), \tag{36}$$

where $K_q$ is the normalization constant satisfying $K_q \to 1$ ($q \to 1$). It should be noticed that the marginal, $\int d^3\mathbf{r}_1\, d^3\mathbf{v}_1\, f_q^{(2)}(\mathbf{r}, \mathbf{r}_1, \mathbf{v}, \mathbf{v}_1, t)$ [$\int d^3\mathbf{r}\, d^3\mathbf{v}\, f_q^{(2)}(\mathbf{r}, \mathbf{r}_1, \mathbf{v}, \mathbf{v}_1, t)$], is radically different from $f(\mathbf{r}, \mathbf{v}, t)$ [$f(\mathbf{r}_1, \mathbf{v}_1, t)$], due to the fact that the marginal is a distribution of a single particle experiencing the influence of another particle through the specific pattern of correlation in Eq. (36), whereas $f$ is a distribution of a single isolated particle.

Correspondingly, Eq. (28) is also generalized to

$$R_q(f, f_1; f', f_1') = K_q\, (f' *_q f_1' - f *_q f_1), \tag{37}$$

which tends to $R(f, f_1; f', f_1')$ in Eq. (28) in the limit $q \to 1$. Actually, it is possible to formally generalize the binary collision to multi-particle collision. The binary collision considered here is nothing but a simplifying assumtion.

Thus, the generalized Boltzmann equation is given by

$$\frac{\partial f}{\partial t} + \mathbf{v} \cdot \nabla f + \frac{\mathbf{F}}{m} \cdot \frac{\partial f}{\partial \mathbf{v}} = \int d\omega \, d^3\mathbf{v} \, \sigma \, V_r \, R_q(f, f_1; f', f_1'). \tag{38}$$

Substituting Eq. (38) into Eq. (24), we have

$$\frac{\partial H_q(\mathbf{r}, t)}{\partial t} = -\frac{1}{1-q} \int d^3\mathbf{v} \, \mathbf{v} \cdot \nabla(f - f^q) - \frac{1}{1-q} \int d^3\mathbf{v} \, \frac{\partial}{\partial \mathbf{v}} \cdot \left[ \frac{\mathbf{F}}{m}(f - f^q) \right]$$

$$+ \frac{q}{1-q} \int d\omega \, d^3\mathbf{v} \, d^3\mathbf{v}_1 \, \sigma \, V_r \, (1/q - f^{q-1}) \, R_q(f, f_1; f', f_1'). \tag{39}$$

The second term on the right-hand side vanishes for $f$ such that $f^q$ ($q > 0$) is integrable. Therefore, Eq. (39) is rewritten as follows:

$$\frac{\partial H_q}{\partial t} + \nabla \cdot \mathbf{j}_q = G_q, \tag{40}$$

where the current, $\mathbf{j}_q$, and the source, $G_q$, are given by

$$\mathbf{j}_q(\mathbf{r}, t) = \frac{1}{1-q} \int d^3\mathbf{v}\, \mathbf{v}\, (f - f^q), \qquad (41)$$

$$G_q(\mathbf{r}, t) = \frac{q}{1-q} \int d\omega\, d^3\mathbf{v}\, d^3\mathbf{v}_1\, \sigma\, V_r\, (1/q - f^{q-1})\, R_q(f, f_1; f', f_1'), \qquad (42)$$

respectively.

Let us evaluate the source term by making use of the standard consideration. $G_q$ should be invariant under the interchange, $\mathbf{v} \leftrightarrow \mathbf{v}_1$, so as are the cross section and $V_r$. Also, the measure is invariant through collision: $d^3\mathbf{v}\, d^3\mathbf{v}_1 = d^3\mathbf{v}'\, d^3\mathbf{v}_1'$. Using the structure of Eq. (37) in Eq. (42) as well as these symmetries, we obtain

$$G_q(\mathbf{r}, t) = -\frac{q}{4(q-1)} \int d\omega\, d^3\mathbf{v}\, d^3\mathbf{v}_1\, \sigma\, V_r$$
$$\times \left[ (f'^{\,q-1} + f_1'^{\,q-1} - 1) - (f^{q-1} + f_1^{q-1} - 1) \right] R_q(f, f_1; f', f_1'). \qquad (43)$$

The integrand on the right-hand side of this equation has the following structure:

$$g_q \equiv \frac{1}{q-1} (x - y) \left\{ [x]_+^{1/(q-1)} - [y]_+^{1/(q-1)} \right\}, \qquad (44)$$

with $x = f'^{\,q-1} + f_1'^{\,q-1} - 1$ and $y = f^{q-1} + f_1^{q-1} - 1$.

Recalling the scheme in Eq. (35), it is clear that $g_q$ above is always nonnegative for $q > 0$. Therefore, we finally conclude that

$$\frac{\partial H_q}{\partial t} + \nabla \cdot \mathbf{j}_q = G_q \leq 0, \tag{45}$$

which generalizes the ordinary *H*-theorem.

Closing this section, let us consider a stationary state, in which the equality, $G_q = 0$, holds. That is,

$$f *_q f_1 = f' *_q f_1'. \tag{46}$$

Using Eq. (31), we have

$$\mathrm{Ln}_q(f) + \mathrm{Ln}_q(f_1) = \mathrm{Ln}_q(f') + \mathrm{Ln}_q(f_1'). \tag{47}$$

This implies that $\mathrm{Ln}_q(f)$ is an additive invariant through collision.

In a special case when the system can approximately be homogeneous without external forces (but still with correlation), the additive invariants in kinematics are only the mass, energy and momentum. Accordingly, $\mathrm{Ln}_q(f)$ is written as the sum of these quantities:

$$\mathrm{Ln}_q(f) = a_0 m + \mathbf{a}_1 \cdot (m\mathbf{v}) + a_2 \frac{1}{2} m \mathbf{v}^2, \tag{48}$$

which leads to

$$f(\mathbf{v}) = N E_q\left(-\lambda (\mathbf{v} - \mathbf{v}_0)^2\right), \tag{49}$$

where $N$, $\lambda$, and $\mathbf{v}_0$ are related to the constants, $a_0$, $\mathbf{a}_1$, and $a_2$, as follows:

$$N = \left[1 + (1 - 1/q)\left(a_0 m - \frac{m\mathbf{a}_1^2}{2a_2}\right)\right]_+^{1/(q-1)}, \qquad (50)$$

$$\lambda = \frac{m a_2}{(1 - 1/q)\left[\dfrac{m\mathbf{a}_1^2}{a_2} - 2a_0 m\right] - 2}, \qquad (51)$$

$$\mathbf{v}_0 = -\frac{\mathbf{a}_1}{a_2}, \qquad (52)$$

provided that the constants have to satisfy the condition: $\lambda > 0$. The distribution in Eq. (49) is normalizable if $q > 1/3$.

Thus, we see that the stationary solution of the generalized Boltzmann equation in Eq. (38) is given in the homogeneous approximation by the $E_q$-distribution, which is a generalization of the Maxwellian distribution centered at $\mathbf{v} = \mathbf{v}_0$.

Finally, we wish to make the following comment. According to the latest development, nonextensive statistical mechanics may be relevant only to discrete systems. Correspondingly, all the above discussions have to be understood in terms of a discrete physical setting of the problem. The generalized Boltzmann equation in Eq. (38) should be interpreted as an approximation of the one on a lattice, for example.

Closing this section, we may mention three earlier works, in which the $H$-theorems are discussed for generalized "entropies" [23] and generalized "relative entropies" [24,25]. However, their relevance to the present work is quite marginal, since they are based on the Markovian master equations, not the Boltzmann-type equations.

## IV. COMMENT ON SHORE-JOHNSON THEOREM

As we have seen above, what to be employed in nonextensive statistical mechanics may be not the $q$-averages but the normal averages. This result is also supported by other independent discussions [9,20]. On the other hand, it is pointed out however in Refs. [11,12] that the theorem of Shore and Johnson [10] prefers the use of the $q$-averages to the normal averages. So, it is necessary to carefully reexamine the theorem from the physical viewpoint. Below, we wish to make a brief comment on this issue.

Shore and Johnson set up the following five axioms:

(i) *Axiom I* (uniqueness): If the same problem is solved twice, then the same answer is expected to result both times.

(ii) *Axiom II* (invariance): The same answer is expected when the same problem is solved in two different coordinate systems, in which the posteriors in the two systems should be related by the coordinate transformation.

(iii) *Axiom III* (system independence): It should not matter whether one accounts for independent information about independent systems separately in terms of their marginal distributions or in terms of the joint distribution.

(iv) *Axiom IV* (subset independence): It should not matter whether one treats independent subsets of the states of the systems in terms of their separate conditional distributions or in terms of the joint distribution.

(v) *Axiom V* (expansibility): In the absence of new information, the prior (i.e., the reference distribution) should not be changed.

Then, they prove the theorem that the relative entropy with the prior $\{r_i\}_{i=1,2,...,W}$ and the posterior $\{p_i\}_{i=1,2,...,W}$ satisfying the above set of axioms has the form

$$I[p\|r] = \sum_{i=1}^{W} p_i\, h(p_i/r_i), \qquad (53)$$

where $h(x)$ is some function. And it is shown in Refs. [11,12] that such $h(x)$ certainly exists, if the *q*-average formalism is employed, whereas the normal-average formalism does not possess such a function. Thus, the Shore-Johnson theorem supports the use of the *q*-average formalism.

Now, the question is if Axioms I-V are acceptable in view of the physical conditions assumed in nonextensive statistical mechanics. To answer it, we need recall the following fact. Nonextensive statistical mechanics is intended for complex systems, in which correlations between elements are strong. In other words, the factorization of the joint distribution as in Eq. (27) cannot be realized. This observation leads to the conclusion that the assumption of system independence in Axiom III is not physically appropriate. It is our opinion that this is why the Shore-Johnson theorem is outside the scope of nonextensive statistical mechanics.

## V. CONCLUSION

We have found that, quite remarkably, kinetic theory is able to specify the correct definition of average to be employed in nonextensive statistical mechanics. We have shown that the normal-average formalism is consistent with the *H*-theorem with the generalized Stosszahlansatz, whereas the *q*-average formalism is not. In particular, we have carefully analyzed the distributions with finite cut-off factors. Accordingly, we have presented an amendment of nonextensive statistical mechanics based on the normal averages. In addition, we have also carefully reexamined the Shore-Johnson theorem, which supports the use of the *q*-average formalism, and have discussed that one of the Shore-Johnson axiom is physically inappropriate for complex systems to be treated within the framework of nonextensive statistical mechanics.

The $E_q$-distribution arising from the normal-average formalism has an intriguing analogy with the property of nonlinear dynamics. The distribution of trajectories of a dynamical system at the edge of chaos and the associated entropy are structurally in parallel with the present ones [26].

## ACKNOWLEDGMENTS

The author would like to thank G. Kaniadakis and Q. A. Wang for discussions and encouragements. This work was supported in part by a Grant-in-Aid for Scientific Research from the Japan Society for the Promotion of Science.

_______________________________________